\documentclass[%
 reprint,
 amsmath,amssymb,
 aps,prl,longbibliography,
]{revtex4-1}

\usepackage{graphicx} 
\usepackage{dcolumn} 
\usepackage{bm} 
\usepackage{xcolor} 

\usepackage{soul} 

\usepackage[none]{hyphenat} 
\usepackage{relsize} 
\usepackage[capitalise]{cleveref} 
\usepackage[shortlabels]{enumitem} 
\setlist[enumerate]{itemsep=0mm} 

\begin{document}
\preprint{APS/123-QED}

\title{Quasi-isodynamic stellarators with low turbulence as fusion reactor candidates}

\author{Alan G. Goodman}
\email{alan.goodman@ipp.mpg.de}
\author{Pavlos Xanthopoulos}
\author{Gabriel G. Plunk}
\author{H\r{a}kan Smith}
\author{Carolin N{\"u}hrenberg}
\author{Craig D. Beidler}
\author{Sophia A. Henneberg}
\author{Gareth Roberg-Clark}
\author{Michael Drevlak}
\author{Per Helander}
\affiliation{Max-Planck-Institut f{\"u}r Plasmaphysik, D-17491 Greifswald, Germany.} 

\date{\today}

\begin{abstract}
The stellarator is a type of fusion energy device that --- if properly designed --- could provide clean, safe, and abundant energy to the grid.
To generate this energy, a stellarator must keep a hot mixture of charged particles (known as a plasma) sufficiently confined by using a fully shaped magnetic field.
If this is achieved, the heat from fusion reactions within the plasma can be harvested as energy.
We present a novel method for designing reactor-relevant stellarator magnetic fields, which combine several key physical properties.
These include plasma stability, excellent confinement of the fast moving particles generated by fusion reactions, and reduction of the turbulence that is known to limit the performance of the most advanced stellarator experiment in the world, Wendelstein 7-X.
\end{abstract}

\maketitle

\section{\label{sec:bg}Introduction}
Fusion has, for decades, been heralded as humanity's ideal energy source, if only it could be realized practically. 
A hypothetical fusion reactor would produce clean, abundant, safe, and energy-dense power \cite{Boozer_2021} through the fusion of deuterium and tritium (DT) atoms.
Fusion is the process that powers the Sun and other stars, meaning that such a reactor would be, in essence, a ``star in a bottle''.
Stars facilitate fusion reactions by virtue of the material that comprises them, a \textit{plasma}, which is held together by the stars' immense gravity.
As we cannot create such strong gravitational fields in an Earth-bound fusion reactor, we must be a bit more creative with how we cage this plasma.

Stellarators are a family of fusion devices that rely on twisted magnetic fields to confine plasmas inside a toroidal vessel.
If a fusion-relevant DT plasma is maintained, fusion reactions will naturally occur between particles within it, generating energy that could be harvested to power an electrical grid.

Such plasmas can be designed through optimization, where the magnetic geometry is tailored to stabilize the plasma and minimize its heat and particle losses.
Several key properties and loss mechanisms --- including bulk stability and the collisional transport of particles --- have already been addressed with the design, construction, and successful operation of existing devices, most notably the world's largest stellarator, Wendelstein 7-X (W7-X) \cite{w7x-grieger}, operating in Greifswald, Germany.
Other aspects, such as adequate fast ion confinement and the reduction of transport caused by plasma turbulence, are areas of active research, as these features --- particularly turbulence --- are outstanding problems in W7-X.

Our tackles these outstanding problems, specifically for a class of stellarator configurations called ``quasi-isodynamic'' (QI), by introducing a method to find such stellarator designs that combine traditional criteria with those needed for a next-generation fusion device.
This includes, for the first time in an optimized QI stellarator, the direct minimization of losses caused by plasma turbulence.

The QI stellarator may be compared with a more conventional and simply-shaped type of fusion device, the \textit{tokamak} \cite{tokamaks}.
Tokamaks keep the particles generated from DT fusion reactions, known as ``fast ions'', contained within their plasmas.
These ions carry large amounts of energy, and thus it is crucial that they remain confined within a fusion reactor.
Stellarators do not confine these particles unless specifically designed to do so.

Tokamaks are able to confine fast ions by driving strong currents in their plasmas.
These currents are essential for tokamak operation, but can also cause severe disruptions and inhibit steady-state operation.
Unlike tokamaks, stellarators do not require externally-driven currents, however they may still be plagued by the so-called \textit{bootstrap} current which spontaneously arises in their plasmas \cite{helander2014,landremanbuller2022}.
This current vanishes if a stellarator is designed to be exactly QI, allowing these stellarators to enjoy ``current-free'' operation \cite{helander2009}.
It has also been shown that some amount of deviation from an exactly QI field can still yield low bootstrap currents \cite{goodman2022, ciemat2023}.

The attractive properties of QI stellarators stem from the motion of ``trapped'' particles, confined to regions where the magnetic field strength $B$ remains below some value $B_*$ (the exact value of which is determined by the direction of the particle's motion).
When a trapped particle encounters a field strength equal to its associated $B_*$, it ``bounces'' back in the direction from which it came.
A conserved quantity of the bouncing particle motion is the \textit{second adiabatic invariant}, defined as
\begin{equation}
    \mathcal{J} = \sqrt{2m\mathcal{H}} \int_\textrm{bounce} \sqrt{1 - \frac{B}{B_*}}\, \textrm{d}l,
\end{equation}
where the ``bounce integral'' is taken between the points along a field line at which $B = B_* = \mathcal{H}/\mu$,
$l$ is the distance along said field line, $\mathcal{H}=mv^2/2$ is the kinetic energy,
$\mu = mv_\perp^2/2B$ is the first adiabatic invariant of the particle,
and $m$ is the particle's mass.

Trapped particles in a QI magnetic field can exhibit the \textit{maximum}-$\mathcal{J}$ \textit{property}, mathematically expressed as ${d}\mathcal{J}/\textrm{d}s < 0$, for a fixed $B_*$.
Here $s$ denotes the normalized toroidal magnetic flux, acting as a radial variable, such that $s=0$ holds at the plasma center and $s=1$ at the plasma boundary.
In other words, in a ``max-$\mathcal{J}$'' field, $\mathcal{J}$ will be largest in the plasma center and smallest at its edge. The max-$\mathcal{J}$ property improves the confinement of fast particles \cite{goodman2022,ciemat2023,Velasco_2023}, while imparting a high degree of magnetohydrodnamics (MHD) stability, related to the vacuum ``magnetic well'' \cite{helander2014,rodriguez2023,rodriguez2023maximumj}. 

Turbulence driven by trapped electrons is relatively benign in experimental campaigns in W7-X \cite{Bozhenkov_2020,pavlos2020}, but turbulence driven by the ion temperature gradient (ITG) is not.
Because, as observed in W7-X \cite{Beurskens_2021,bahner_2021,Carralero_2021}, ITG turbulence is expected to limit modern stellarator performance, next-generation stellarator reactor designs would benefit greatly if energy losses driven by ITG turbulence could be made significantly lower than in W7-X.

Our ITG turbulence minimization strategy is based on two observations of how this turbulence behaves.
First, ITG turbulence is generally excited in regions where magnetic curvature aligns with the plasma temperature gradient, which we call ``bad'' curvature.
Second, the {intensity} of this turbulence depends on the compression of neighboring magnetic flux surfaces at these locations \cite{stroteich2022}. 
Intuitively, this follows from the expression for the temperature gradient:
$
    |\nabla T| = T'(s) |\nabla s|.
$
For a fixed plasma profile $T(s)$, lower flux surface compression (\textit{i.e.,} smaller $|\nabla s|$) leads to lower ITG turbulence.

\begin{figure}[ht]
    \includegraphics[width=0.75\columnwidth]{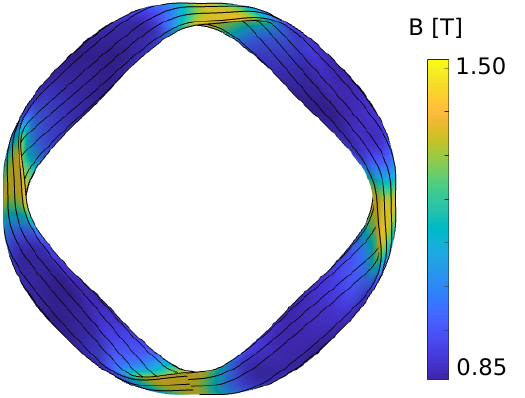}
    \caption{The plasma boundary of a SQuID configuration. Some field lines are shown in black.
    }
    \label{fig:SQuID_3D}
\end{figure}

To date, no MHD-stable, max-$\mathcal{J}$, QI stellarators have been designed for which ITG turbulence has been specifically minimized.
Here, we outline a novel approach to designing such stellarators, to which we give the moniker \textit{S}table \textit{Qu}asi-\textit{I}sodynamic \textit{D}esigns (SQuIDs).
The SQuID presented in this work is shown in \cref{fig:SQuID_3D}.

We compare this SQuID with another QI configuration found through the same approach, but without ITG turbulence minimization, allowing us to explore trade-offs regarding the desired physics properties.
For reference, these new configurations are also compared against W7-X.
We find that this SQuID addresses all the criteria we set out to achieve: its QI quality grants excellent fast ion confinement, good confinement of collisional thermal particles, and small toroidal currents.
It is MHD stable, owing to the max-$\mathcal{J}$ property, which --- along with our explicit turbulence reduction scheme --- yields low levels of turbulence.

\section{\label{sec:methods}Optimization Method}
Given the shape of a stellarator plasma's boundary, along with its pressure and current profiles, one can calculate its equilibrium magnetic field \cite{kruskal1958}.
In other words, given these profiles (which are generally fixed during optimization), a boundary shape is sufficient information for calculating many properties of the magnetic field throughout that plasma. The plasma's boundary surface shape can be given by a Fourier series:
\begin{equation}
    \begin{gathered}
        R(\theta,\phi) = \sum_{m,n} r_{mn}\cos(m\vartheta - n_\textrm{fp}n\phi) \\
        Z(\theta,\phi) = \sum_{m,n} z_{mn}\sin(m\vartheta - n_\textrm{fp}n\phi) 
    \end{gathered}
\end{equation}
where $\phi$ is the azimuthal toroidal angle, $\vartheta$ is a poloidal angle, and $n_\textrm{fp}$ is the number of field periods \cite{VMEC,henneberg_helander_drevlak_2021,landremanpaul2022}.
To optimize the plasma boundary, $r_{mn}$ and $z_{mn}$ are varied, such that various ``target functions'', listed below, are minimized. 
These optimizations were done using the \texttt{simsopt} suite \cite{simsopt1} and the equilibrium code \texttt{VMEC} \cite{VMEC}, following the general methodology of \citet{goodman2022}.

\subsection{QI target}

\begin{figure}[ht]
    \includegraphics[width=0.7\columnwidth]{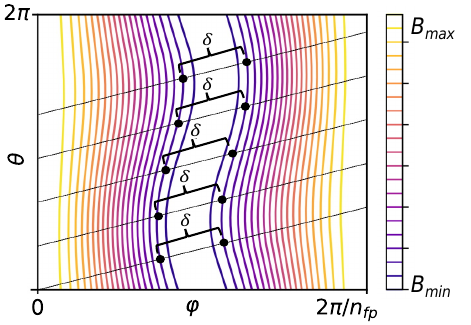}
    \caption{$B$ contours of a perfectly QI field on a flux surface, in Boozer coordinates. Magnetic field lines are shown in black.}
    \label{fig:QI_example}
\end{figure}

Defined intuitively, the magnetic field on a flux surface is QI if, and only if, it satisfies three conditions \cite{Cary1997,goodman2022} when plotted in Boozer coordinates \cite{Boozer1981} ($\theta,\varphi$): 
\begin{enumerate}[(1)]
    \item All curves of constant magnetic field close poloidally (but not toroidally) on all flux surfaces, (\textit{i.e.,} from bottom to top in \cref{fig:QI_example}),
    \item The contours of maximum field strength must be straight lines at toroidal angle $\varphi=0$ and $2\pi/n_\textrm{fp}$, with $n_\textrm{fp}$ the number of identical field periods of the stellarator, and
    \item The \textit{bounce distance} ($\delta$) along a field line between consecutive points with $B=B_*$ (with $B<B_*$ between the points) must be the same along each field line.
\end{enumerate}

\begin{figure}[ht]
    \includegraphics[width=0.8\columnwidth]{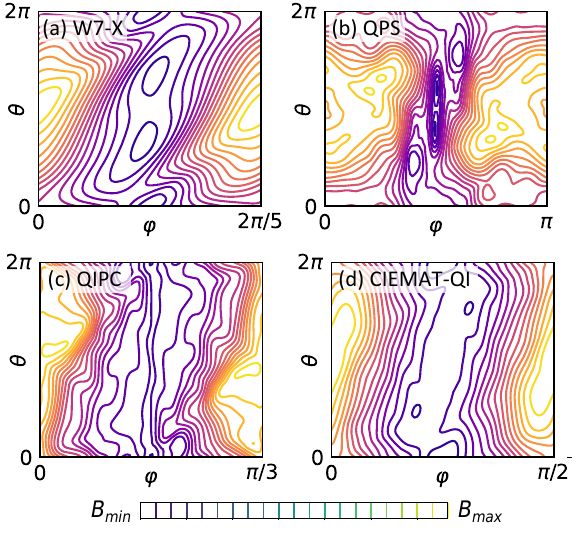}
    \caption{$B$ contours near the plasma boundary, in Boozer coordinates, for various QI configurations (coil ripple is not included): (a) W7-X, (b) QPS \cite{qps}, (c) QIPC \cite{qipc,qipc:nf:2006}, and (d) CIEMAT-QI \cite{ciemat2023}.
    }
    \label{fig:QI_configs}
\end{figure}

An artificial, perfectly QI field is shown in \cref{fig:QI_example}.
In \cref{fig:QI_configs}, we show the magnetic field for a set of actual configurations that are approximately QI.

%

In a QI field, $\partial\mathcal{J}/\partial\alpha|_{s,B_*} = 0$ holds for any $\{\alpha,B_*\}$, where $\alpha$ is the field line label, defined by $\mathbf{B} = \nabla \psi \times \nabla \alpha$.

To obtain QI fields through numerical optimization, we penalize unwanted defects of an input equilibrium field $B_I(s,\theta,\varphi)$.
To do this, we first take a similar approach to the \textit{Squash} and \textit{Stretch} described in \citet{goodman2022}, which transforms $B_I$ into a constructed field $B_C(s,\theta,\varphi)$ with no local trapping wells or transitioning particles.
With $\lambda=1/B_*$, we then define the following quantities:
\begin{eqnarray}
    \tilde{\mathcal{J}}_I(\lambda,\alpha) 
    &=&
    \int_{\varphi_1}^{\varphi_2} 
    \textrm{sign}\left(1 - \lambda B_I\right) 
    \sqrt{\left| 1 - \lambda B_I \right|}
    \,\,\textrm{d}l_I, 
    \\
    \tilde{\mathcal{J}}_C(\lambda,\alpha) 
    &=& 
    \int_{\varphi_1}^{\varphi_2}  
    \sqrt{ 1 - \lambda B_C }
    \,\,\textrm{d}l_I,
\end{eqnarray}
where 
$
\textrm{d}l_I = B_I\,\textrm{d}\varphi/(\boldsymbol{B}_I\cdot\nabla \varphi)
$ and $(\varphi_1,\varphi_2)$ are the bounce points in $B_C$.
The QI penalty function, $f_\textrm{QI}$, is
\begin{equation}
    f_\textrm{QI}(s) \propto 
    \mathlarger{\mathlarger{\sum}}_{\lambda,i,j}
    \left(
        \frac{\tilde{\mathcal{J}}_I(\lambda,\alpha_i) - \tilde{\mathcal{J}}_C(\lambda,\alpha_j)}
        {\langle \tilde{\mathcal{J}}_I + \tilde{\mathcal{J}}_C \rangle} 
    \right)^2,
    \label{eq:fqi}
\end{equation}
where $\langle\cdot\rangle$ denotes an average of all calculated values.
Note that $f_\textrm{QI}=0$ if, and only if, $B_I$ is perfectly QI.  This target is used to improve what we will refer to as ``QI quality'', but the ultimate quantification of such quality must be deferred to the evaluation of the key physics quantities, {\em i.e.}, bootstrap current, fast particle confinement, and neoclassical effective ripple.

\subsection{Max-$\mathcal{J}$ target}
Using the definitions above, we can design a target function that favors max-$\mathcal{J}$ fields. First, we evaluate $\tilde{\mathcal{J}}_C$ on multiple flux surfaces, and define
\begin{multline}
    \partial_s \tilde{\mathcal{J}}(s,\lambda,\alpha_i,\alpha_j) = \\
    \frac{1}{\Delta s} 
    \frac{\tilde{\mathcal{J}}_C(s + \Delta s, \lambda, \alpha_i) - \tilde{\mathcal{J}}_C(s, \lambda, \alpha_j)}{\langle \tilde{\mathcal{J}}_C(s + \Delta s) + \tilde{\mathcal{J}}_C(s) \rangle}.
\end{multline}
By definition, a max-$\mathcal{J}$ field satisfies the condition $\partial_s \mathcal{J} < 0$. We also seek a large $|\partial_s\mathcal{J}|$, in order to improve fast-particle confinement \cite{ciemat2023,goodman2022} and mitigate the turbulence driven by trapped particles \cite{goodman2022,proll2022,helander2012}. 
We, hence, define the max-$\mathcal{J}$ target as
\begin{equation}
    f_{\textrm{max}\mathcal{J}} \propto 
    \mathlarger{\sum}_{\lambda,i}
    M_t\left( 
    \langle \partial_s \tilde{\mathcal{J}}(s,\lambda,\alpha_i,\alpha_j)\rangle_{\alpha_j},
    T_\mathcal{J}
    \right)
    \label{eq:fmJ}
\end{equation}
with $M_t(X,X_t) \equiv \max(0,X-X_t)^2$, and $\langle\cdot\rangle_{\alpha_j}$ being an average over $\alpha_j$.
Based on a numerical study of the values of $\partial_s \tilde{\mathcal{J}}$ of well-confined particles in highly optimised QI stellarators (similar to those presented in \citet{goodman2022}), we chose a target radial derivative $T_\mathcal{J}=-0.06$.
A more negative $T_\mathcal{J}$ targets ``more max-$\mathcal{J}$'' fields.

\subsection{ITG target}
Pioneering efforts in turbulence optimization have devised targets for reducing ITG turbulence, for instance, by reducing the curvature of the magnetic field \cite{Mynick_2010,Mynick_2011,Mynick_2014,Xanthopoulos2014} or by exploiting the mode properties close to the threshold for the onset of the instability \cite{robergclark2023}. In this work, we follow a related approach, which we find compatible with the simultaneous optimization for QI quality of the magnetic field.
We use a target to minimize $|\nabla s|$ in regions of ``bad'' curvature (defined by $\tilde{\kappa}<0$, where
$
\tilde{\kappa} = \mathbf{B} \times \boldsymbol{\kappa} \cdot \nabla\alpha
$) as ITG turbulence tends to peak in regions of bad curvature. 
An appeal of this approach is its simplicity and physical transparency (although newer approaches that use turbulence simulations as optimization criteria \cite{kim2023optimization} also show promise).

We begin by calculating $\tilde{\kappa}$ and $\nabla s$ on a fine grid on a flux surface, and construct $\xi = (a_\textrm{min}\Theta(-\tilde{\kappa}) |\nabla s|)^2$, where $\Theta$ is the Heaviside step function and $a_\textrm{min}$ is the plasma boundary's effective minor radius, following the convention in the \texttt{VMEC} code. 
We then define $\xi_{95}$ as the value of $\xi$ below which 95\% of the calculated values lie, and construct the target function $f_{\nabla s}$ to be
%
\begin{equation}
    f_{\nabla s} \propto 
    \int_0^{2\pi} \textrm{d}\theta 
    \int_0^{2\pi/n_\textrm{fp}} \textrm{d}\phi \,\,
    \xi(\theta,\phi)\Theta(\xi_{95}-\xi(\theta,\phi)).
\end{equation}
This approach ensures that the number of points with $\tilde{\kappa}<0$ will not impact the target function output. 

\subsection{Other targets}
To control other properties of interest, we include a few generic terms in our target function.
We use the function $f_\Delta\propto M_t(\Delta,\Delta_t)$ to control the mirror ratio $\Delta = (\mathcal{B}_\textrm{max} - \mathcal{B}_\textrm{min})/(\mathcal{B}_\textrm{max} + \mathcal{B}_\textrm{min})|_{s=0}$, where $\mathcal{B}_\textrm{min}$ ($\mathcal{B}_\textrm{max}$) is the minimum (maximum) $B$ on the flux surface.
We also introduce $f_A\propto M_t(A,A_t)$ and $f_\beta\propto M_t(\beta,\beta_t)$ to limit the aspect ratio $A$ and the average normalized plasma pressure 
$
    \beta= \int_0^1 \textrm{d}s\left\langle 2\mu_0p(s)/B(s)^2 \right\rangle_s
$ 
respectively (both calculated by \texttt{VMEC}).
We define $f_\iota \propto M_t(|\iota(s=0)-\iota_\textrm{ax}|,0.01) + M_t(|\iota(s=1)-\iota_\textrm{edge}|,0.01)$ which sets $\iota$ on axis ($\iota_\text{ax}$) and at the plasma boundary ($\iota_\text{edge}$).

The target function used for the SQuID optimization is thus
$
    f_\textrm{SQuID} = f_\textrm{QI} + w_{\nabla s} f_{\nabla s} + f_{\textrm{max}\mathcal{J}} + f_\Delta + f_A + f_\beta + f_\iota,
$
where $f_\textrm{QI}$ and $f_{\textrm{max}\mathcal{J}}$ are targeted on $s\in(0,1)$, and $f_{\nabla s}$ is targeted on $s\in[0.1,0.5]$. 
Note that every component of $f_\textrm{SQuID}$ is an inequality constraint, except for $f_\textrm{QI}$ and $f_{\nabla s}$.
Hence, during optimization, we set the weightings for these inequality constraints to be arbitrarily high, and adjust only the relative weight, $w_{\nabla s}$, between $f_\textrm{QI}$ and $f_{\nabla s}$. 
For the SQuID, $w_{\nabla s}$ is set to the largest value that maintained other properties of interest in the configuration.
We compare the SQuID to another QI configuration, optimized in the same way, but with $w_{\nabla s}$ set to zero.

\section{Results}\label{sec:results}
\begin{figure}[ht]\includegraphics[width=0.75\columnwidth]{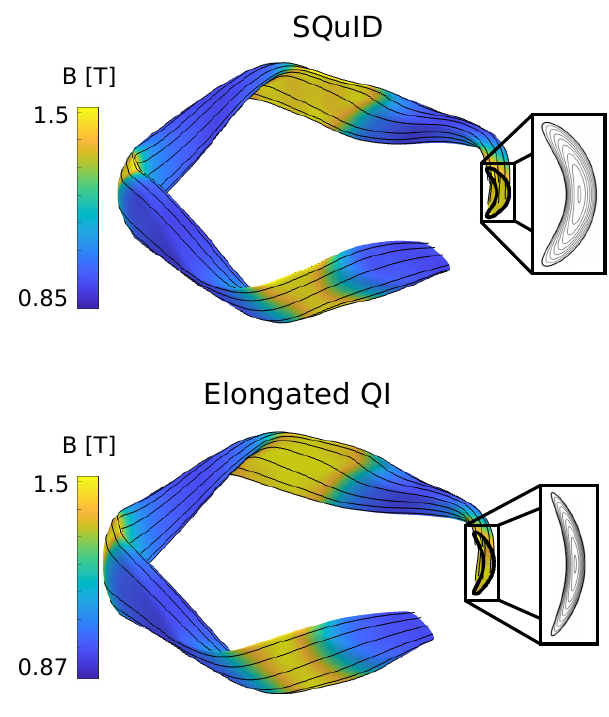}
    \caption{Plasma boundaries of the SQuID (top) and elongated QI configuration (bottom).
    Half a field-period is removed to highlight the most compressed plasma cross-sections.
    }
    \label{fig:3Dplots}
\end{figure}

The configurations shown in \cref{fig:3Dplots} were optimized starting from the W7-X standard configuration, imposing $n_\textrm{fp}=4$. 
Both configurations have the same aspect ratio ($A\simeq10$), mirror ratio ($\Delta\simeq0.25$), pressure profile ($p(s)\propto1-s$), normalized plasma pressure ($\beta=2\%$), and rotational transform ($\iota \simeq [0.80, 0.95]$).
Optimizations were also performed for $\iota \simeq [0.86, 0.95]$ to avoid the 4/5 rational surface, with nearly identical results.

\begin{figure}[ht]
    \includegraphics[width=\columnwidth]{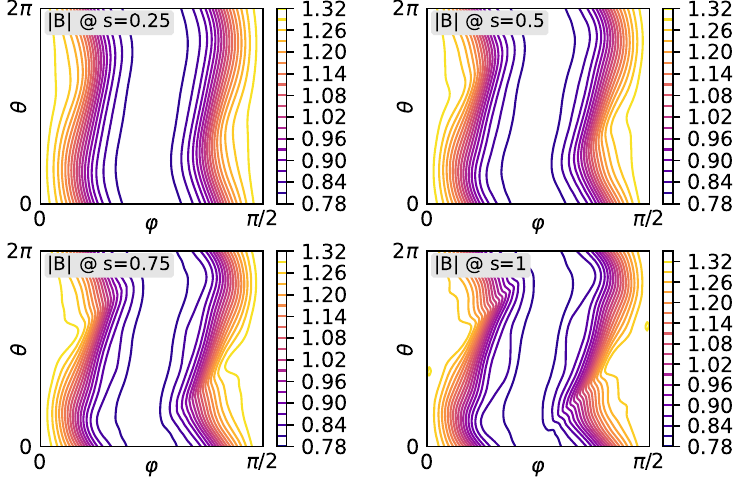}
    \caption{$B$ contours, in Boozer coordinates, on four flux surfaces for the elongated QI configuration.}
    \label{fig:nautB}
\end{figure}

The configuration for which $w_{\nabla s}=0$ has excellent QI quality (see \cref{fig:nautB}), confirming the effectiveness of our QI target.
Its flux surfaces, however, appear extremely elongated (see \cref{fig:3Dplots}), a feature associated with large $|\nabla s|$ and, thus, strong ITG drive.
We dub this configuration ``elongated QI''.

\begin{figure}[ht]
    \includegraphics[width=\columnwidth]{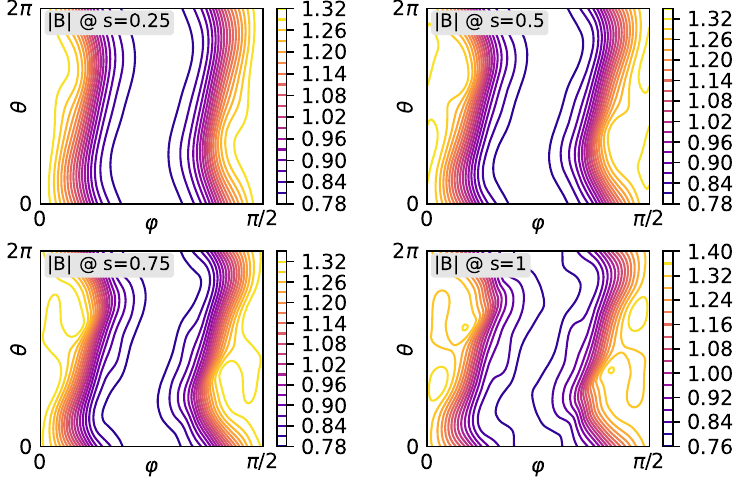}
    \caption{$B$ contours, in Boozer coordinates, on four flux surfaces for the SQuID.}
    \label{fig:vampB}
\end{figure}

The SQuID, on the other hand, is less elongated (see \cref{fig:3Dplots}), but retains good QI quality (see \cref{fig:vampB}) and enjoys all the benefits thereof.
Its maximum value of $a_\textrm{min}|\nabla s|$ in a bad curvature region is $\sim$2.3, compared to $\sim$6.5 in the elongated QI configuration.
This flux compression reduction at the cost of a roughly $1.6\times$ increase in $f_\textrm{QI}$ over the elongated QI configuration.

\section{Analysis}

\subsection{Fast particle confinement}\label{subsec:fastions}
To evaluate the confinement of fusion-born alpha particle orbits, we used the code \texttt{SIMPLE} \cite{simple} to follow the trajectories of 5000 alpha particles for 0.2 seconds (the typical collisional slowing down time in a reactor), with uniformly distributed pitch angle and 3.5 MeV of kinetic energy, isotropically (in real-space) on the flux surface $s=0.25$.
As is common practice, we scaled the mean on-axis field to $B_{00}=5.7\textrm{ T}$, the same as the ARIES-CS stellarator \cite{aries-cs}.

\begin{figure}[ht]
    \includegraphics[width=0.7\columnwidth]{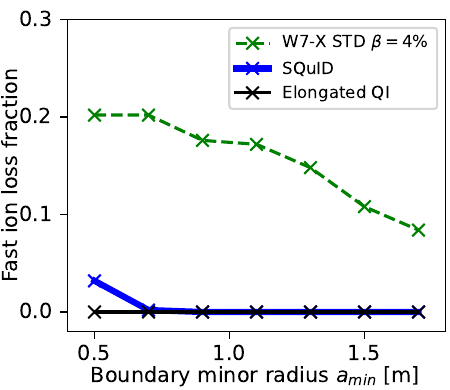}
    \caption{Alpha particle losses for the W7-X STD and the new QI configurations on $s=0.25$, once the boundary is scaled to various minor radii.
    }
    \label{fig:Losses}
\end{figure}

The minor radius of the torus, $a_\textrm{min}$, plays a key role in fast-ion confinement, and at typical reactor scales ($a_\textrm{min}\sim1.7\textrm{ m}$ \cite{landremanpaul2022}) we calculated zero fast ion losses in both QI configurations. Hence, it is predicted that, in a reactor scenario, these configurations would enjoy excellent ion confinement. However, to account for possible variation in actual reactor designs, we repeated the calculation for various values of $a_\textrm{min}$, with the corresponding fast ion loss fractions shown in \cref{fig:Losses}.

It is considered acceptable for a reactor to lose $\sim1\%$ of fast particles generated at $s=0.25$ \cite{lion2023}, which is satisfied for $a_\textrm{min}\gtrsim0.5\textrm{ meters}$ for the elongated QI configuration, and $a_\textrm{min}\gtrsim0.6\textrm{ meters}$ for the SQuID.
For larger $a_\textrm{min}$, both QI configurations have zero losses, indicating that both have reactor-relevant fast ion confinement.

\subsection{Neoclassical transport}\label{subsec:neoclassical}
\begin{figure}[ht]
    \includegraphics[width=0.75\columnwidth]{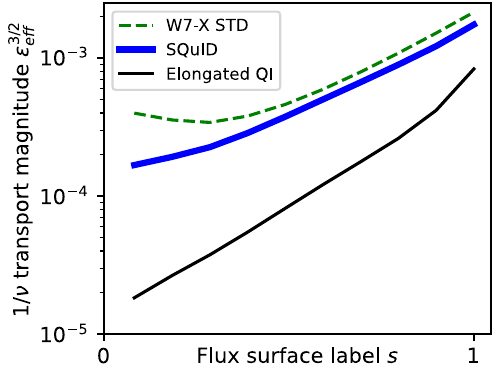}
    \caption{Transport magnitude $\varepsilon_\textrm{eff}^{3/2}$ over the plasma radius, for the W7-X standard and the QI configurations.
    }
    \label{fig:EEs}
\end{figure}

While fast ions hardly interact with thermal particles in a confined plasma, such thermal particles ``collide'' among themselves, causing losses due to neoclassical transport.
In the so-called $1/\nu$ \textit{regime} (here, $\nu$ is the particle collision frequency), weakly-collisional trapped particles may drift out of the plasma. 
These transport fluxes are proportional to the effective ripple (in a conventional stellarator it coincides with the helical ripple) $\varepsilon_\textrm{eff}^{3/2}$ \cite{drevlak2003,nemov1999}, which must be sufficiently small in a fusion reactor.
We used the \texttt{NEO} code \cite{nemov1999} to calculate $\varepsilon_\textrm{eff}^{3/2}$ for the elongated QI configuration and the SQuID, along with the W7-X standard (STD) configuration, shown in \cref{fig:EEs}.


Both new configurations have lower $\varepsilon_\textrm{eff}^{3/2}$ than W7-X STD, which has $\varepsilon_\textrm{eff}^{3/2}$ within an acceptable range for a reactor \cite{Beidler_2011}, meaning that this quantity is also within a reactor-relevant range for the new configurations.

\subsection{Max-$\mathcal{J}$ property}\label{subsec:maxJ}
\begin{figure}[ht]
    \includegraphics[width=\columnwidth]{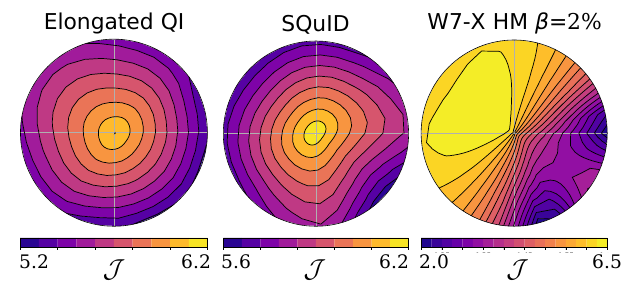}
    \caption{Contours of $\mathcal{J}$, for the 30\% most deeply trapped particles for the new QI configurations and W7-X HM. The radial coordinate is the flux-surface label $s$ and the polar coordinate is the poloidal angle $\theta$. At $\beta=2\%$, the max-$\mathcal{J}$ property is fulfilled for the elongated QI and the SQuID, as $\mathcal{J}$ is largest in the center and decreases towards the plasma radius.}
    \label{fig:PolarJ}
\end{figure}

In \cref{fig:PolarJ}, we display $\mathcal{J}$ contours along the plasma radius, for the elongated QI configuration, the SQuID, and the W7-X ``high-mirror'' (HM) configuration (one of the most max-$\mathcal{J}$ variants of the device).
For all equilibria, the volume-averaged plasma pressure is $\beta=2\%$. We find that, for the 30\% most deeply trapped particles, {\it i.e.} $(B_*-B_{\rm min})/(B_{\rm max} - B_{\rm min}) =0.3$, the QI configurations are max-$\mathcal{J}$, as $\mathcal{J}$ is a decreasing function of the plasma radius.
Similar results were found for all other particles (not shown here).

It should be noted that, if $\beta$ is increased, plasma equilibria become even more max-$\mathcal{J}$.

\subsection{Bootstrap current}\label{subsec:bootstrap}
\begin{figure}[ht]
    \includegraphics[width=\columnwidth]{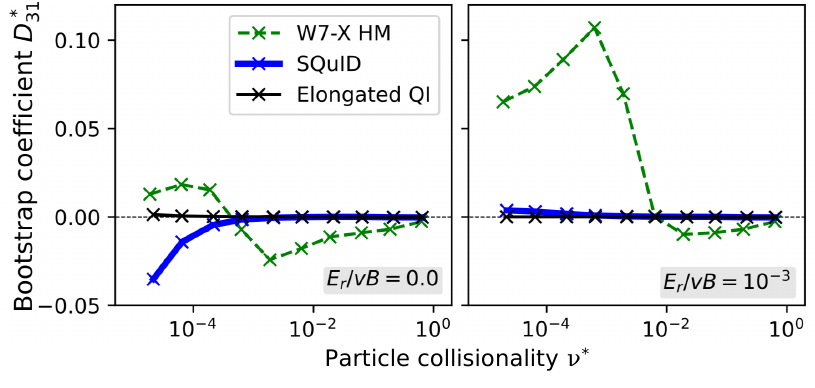}
    \caption{Mono-energetic bootstrap current coefficient $D_{31}^*$, for W7-X HM and the new QI configurations ($s=0.25$), as a function of particle collisionality $\nu^*$, for two values of the radial electric field $E_r/v B$.}
    \label{fig:D31s}
\end{figure}

The toroidal bootstrap current appearing in fusion devices can cause significant problems with both confinement and stability, and interferes with island divertor operation.
It is therefore desirable to minimize this current. Theoretically, in a perfect QI configuration, the bootstrap current would vanish at low collisionality \cite{helander2009,helander2011}.
A well-designed QI target can thus be used to obtain a small bootstrap current \cite{goodman2022}, which is the approach followed here.


We therefore present, in \cref{fig:D31s}, the normalized {\it mono-energetic bootstrap current coefficient} \cite{Beidler_2011} for both new configurations in comparison with the W7-X HM at $s=0.25$, which has the lowest bootstrap current among all reference configurations, and was therefore chosen to make the comparison as stringent as possible.
Here, $\nu^\star = R_{maj}\nu /(\iota v)$ is the mono-energetic collisionality, where $R_{maj}$ is the major radius of the torus and $v$ is the particle velocity.
Two values of normalized radial electric field have been chosen, as they are indicative of values of relevance for electrons ($E_r/vB=0$) and for ions ($E_r/vB=10^{-3}$).

If a configuration has $|D_{31}^\star| < 0.01$ over the entire range of $\nu^\star$ and $E_r/vB$ values of relevance, negligible values of the bootstrap current density are assured.
The elongated QI clearly fulfills this requirement, while the SQuID has $D_{31}^\star$ values somewhat outside of this interval at the smallest values of $\nu^\star$ for $E_r/vB=0$.
We have therefore also calculated profiles of the bootstrap current density in SQuID, for a selection of plasma profiles that would produce 600~MW of $\alpha$-particle heating in a reactor.
In none of these cases is the bootstrap current density large enough to adversely affect the plasma equilibrium, and the total current enclosed within the plasma always remains below 10~kA, so that performance of the island divertor
would also be unaffected.

From these results we conclude that both new configurations have reactor-relevant bootstrap currents.




\subsection{ITG turbulence}\label{subsection:itg}
To evaluate the reduction of ITG-driven turbulence, we perform gyrokinetic \cite{Brizard,Garbet_2010} simulations.
These use a thin ``flux-tube'' domain \cite{Beer}, which follows a magnetic field line for one poloidal turn on the flux surfaces $s=0.09$ (core) or $s=0.5$ (plasma periphery).
The tubes are positioned such that they cross the region with $\theta=\phi=0$, where ITG turbulence fluctuations are generally excited.
Moreover, we neglect gradients that do not drive ITG turbulence (\textit{i.e.,} in density and electron temperature) and also collisions.
The simulations were conducted assuming either vacuum or $\beta=2\%$ conditions, although in both cases the ITG turbulence is assumed electrostatic.
This choice implies the strongest turbulence response, avoiding its reduction by electromagnetic effects \cite{Pueschel_2008,Zocco_2015} in QI configurations showing a small Shafranov shift.

\subsubsection{Growth rates}
\begin{figure}[ht]
    \includegraphics[width=0.7\columnwidth]{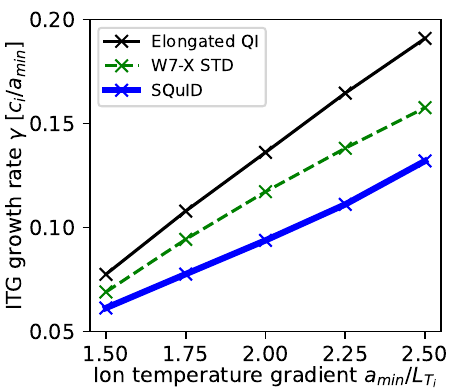}
    \caption{ITG growth rates for W7-X standard and the QI configurations with $\beta=0$ as a function of the ion temperature gradient $a_\textrm{min}/L_{T_i}$.
    }
    \label{fig:gammas}
\end{figure}

Before investigating the turbulence itself, we first study the linear nature of the ITG instability, which --- although not by itself a reliable predictor of the turbulence intensity --- serves as a measure of the turbulence drive.
For this, using the GENE code \cite{GENE}, we calculate the \textit{growth rate} $\gamma$ maximized over the wave number $k_y$ of the instability as a function of the normalized temperature gradient ($a_\text{min}/L_{T_i}=-\frac{a_\text{min}}{T_i}\frac{dT_i}{dr}$, where $T_i$ is the ion temperature and $r = a_\text{min} s^{1/2}$ is the radial coordinate), for the two new configurations and W7-X STD, under vacuum conditions ($\beta=0$).
Here, the electrons are treated in the ``adiabatic'' limit, namely assuming a Boltzmann response and, in all simulations, collisions have been neglected.
This setup, despite its simplicity, already captures the impact of the optimization, as evidenced by the outcome shown in \cref{fig:gammas}.

We observe lower growth rates over the entire range of gradients for the ITG-optimized SQuID, as compared to the elongated QI configuration, which confirms the effectiveness of the target function $f_{\nabla s}$ in reducing the ITG drive.
Moreover, the SQuID is more stable than the W7-X standard configuration, while the higher growth rates of the elongated QI configuration reflect the compressed flux surfaces depicted in Fig.~\ref{fig:3Dplots}.


\subsubsection{Ion heat fluxes}
\begin{figure}[ht]
    \includegraphics[width=1\columnwidth]{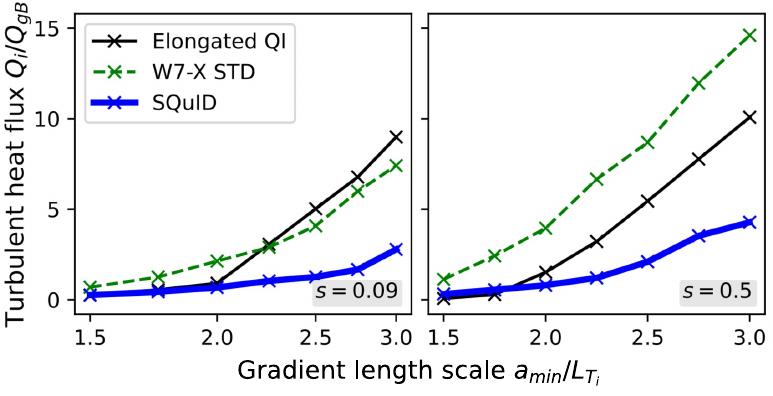}
    \caption{Ion heat fluxes caused by ITG-driven turbulence, for the W7-X standard and the QI configurations (for $\beta=0$), over the ion temperature gradient, on the surfaces $s=0.09$ (left) and $s=0.5$ (right).
    }
    \label{fig:Qi_adiabatic}
\end{figure}

As an overall measure of the severity of turbulence, we calculate the radial ion heat flux (normalized to gyro-Bohm units) $Q_i/Q_{gB}$, while varying the normalized temperature gradient $a_\text{min}/L_{T_i}$. 
In \cref{fig:Qi_adiabatic}, we present simulation results for the two new configurations and the standard W7-X configuration, all at $\beta=0$, using the nonlinear gyrokinetic code GX \cite{mandell2020,mandell2022gx}.

\begin{figure}[ht]
    \includegraphics[width=0.7\columnwidth]{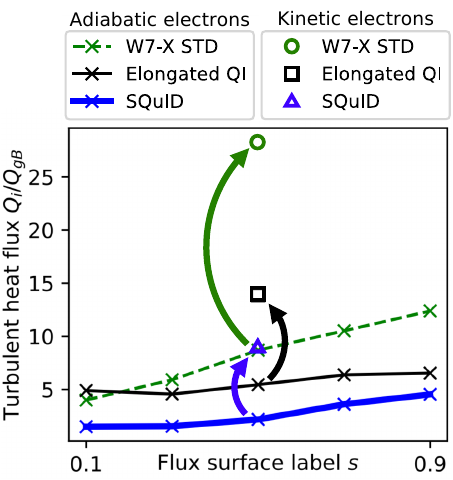}
    \caption{Ion heat fluxes caused by ITG-driven turbulence, for W7-X STD and the new configurations, with adiabatic electrons (lines) and kinetic electrons (symbols) for $a_\textrm{min}/L_{T_i}=2.5$ (see text for details).
    }
    \label{fig:Qi_ak}
\end{figure}

Consistent with the linear picture, the SQuID boasts the lowest heat fluxes over the entire range of gradients among all configurations.
Interestingly, the heat fluxes for the elongated QI configuration are lower than these for the W7-X standard configuration, at least at the outer radial position $s=0.5$.
This seems to contradict the linear observations, suggesting that nonlinear effects of the magnetic geometry are significant.
As explained later, \textit{zonal flows} \cite{Diamond_2005} seem to be an important factor for understanding this behavior.
Overall, the better performance for the SQuID persists for both surfaces and, in actuality, over the entire plasma radius (for a fixed gradient $a_\text{min}/L_{T_i}=2.5$), as shown in \cref{fig:Qi_ak}.

We next address the effect of the  max-$\mathcal{J}$ property on ITG turbulence, which in accordance with \cref{fig:PolarJ}, call for plasmas with $\beta=2\%$, instead of vacuum fields.
The electron dynamics must now also be simulated kinetically to capture the effects of trapped electrons. 

In \cref{fig:Qi_ak}, we include GENE simulation results of these computationally intensive simulations with an additional data point for each of the three configurations, choosing $s=0.5$ and $a_\text{min}/L_{T_i}=2.5$.
In all configurations, the inclusion of kinetic electron physics results in an increase in ion heat flux, compared to the adiabatic electron case. 
This increase for the QI configurations is small as compared to W7-X, presumably due to the max-$\mathcal{J}$ property \cite{proll2022, costello-plunk-helander-arxiv-2024}.
Both new configurations have significantly lower ITG heat flux than W7-X, with the SQuID being the clear winner overall.



\subsubsection{Zonal flows}
\begin{figure}[ht]
    \includegraphics[width=0.7\columnwidth]{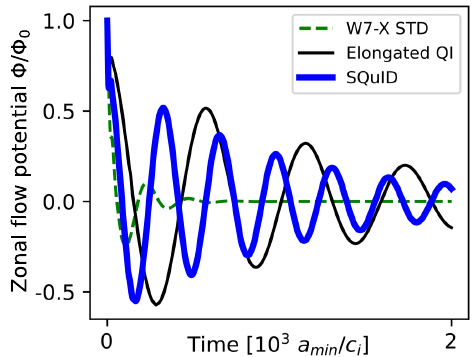}
    \caption{Linear response of zonal flows for the new configurations and W7-X STD.}
    \label{fig:zfs}
\end{figure}

Amidst the chaotic motion of ITG turbulence, zonal flows can evolve as self-organized plasma $\mathbf{E}\times \mathbf{B}$ flows, spanning the entire toroidal magnetic surface and directed mostly poloidally, {\em i.e.} the short way around. These flows can lead to the ``shearing'' of turbulent eddies \cite{hahm_shear}, thus reducing ITG-driven turbulence.
The effect of zonal flows in stellarators, which has been verified in analytical theory as well as experimentally \cite{Helander_zf,Monreal_2016,pavlos_zf,plunk_distinct, tj2}, has characteristics peculiar to stellarator geometry, notably the existence of slowly decaying oscillations \cite{mishchenko2008}.
These oscillations exhibit long characteristic timescales in well-optimized stellarators, much longer than that of the geodesic acoustic modes (GAMs) that are observed in tokamaks.  Slow oscillations are visible in \cref{fig:zfs}, which stand out from the GAMs as the dominant contribution to the frequency spectrum.

It is thought that the decay rate of these oscillations, which largely depends on the specific magnetic geometry, controls the levels of turbulence \cite{Masanori_zf},
as slower-decaying oscillations contribute to turbulence mitigation.

The slow decay rate of the SQuID's zonal flows, in comparison to that of W7-X STD, is striking. Although the detailed mechanism of zonal flow enhancement in the SQuID (and other QI stellarators \cite{goodman2022}, including the elongated QI configuration) warrants further study \cite{plunk-residual-arxiv-2023}, we argue that an improved QI quality of the configuration is related to larger amplitude and slower decay rate of the oscillations.

By performing nonlinear simulations with zonal flows artificially set to zero, we have verified that zonal flows significantly reduce the turbulent heat flux in the elongated QI configuration, {\em e.g.} by a factor of $2.5\times$ for $a/L_{T_i}=2.5$, and contribute to the turbulence reduction in the SQuID as well.

\subsection{MHD stability}\label{subsec:mhd}
The SQuID configuration shows good ideal MHD stability properties,
while the elongated QI case could be considered excellent in this regard.
We evaluated Mercier's criterion \cite{mercier:nf:1962}, a standard measure of MHD stability, using a formulation in magnetic coordinates \cite{jun:tfp:1987}, and found a stability limit of volume-averaged $\beta$ around $2.5\%$ for the SQuID.
The elongated QI configuration appears more stable, with a $\beta$ limit around $8.5\%$. 
This is similar to the MHD stability properties of the QIPC design, which has a stability limit of around $\beta=8\%$ \cite{qipc:nf:2006}, and better than W7-X, which has this limit around $\beta=5\%$ \cite{grieger:iaea:1990}.


\section{Discussion \& outlook}
This work introduces a novel family of QI stellarators, termed ``SQuIDs'' found using a new numerical optimization method, which is capable of addressing a wealth of physics design criteria.
Indeed, for the first time, net-current-free, stable plasmas with excellent particle confinement and low turbulence are attainable, overcoming the main outstanding hurdles on the path towards reactor-relevant equilibria.

These SQuID equilibria are MHD stable, essentially as a result of imposing the maximum-$\mathcal{J}$ property.
The latter is expected to stabilize turbulence caused by the trapped electron mode (TEM) \cite{Rosenbluth1968,helander2012,proll2022,mackenbach2022}.
The optimization was aimed at reducing the ITG turbulent heat flux to levels well below that of the W7-X stellarator, which set the performance limit of the experiment in certain plasma heating scenarios. The imposed ITG target is also expected to control the instability driven by the electron temperature gradient (ETG), due to the dual nature of ETG and ITG modes \cite{GENE}. In addition, ETG turbulence is expected to be of minor concern in reactor conditions, due to the effective coupling of ion and electron temperatures at high plasma densities \cite{ETGturbulence}.
Finally, the presented SQuID was found to enjoy small collisional thermal transport, bootstrap current, and fast-ion losses.

\begin{figure}[ht]
    \includegraphics[width=0.8\columnwidth]{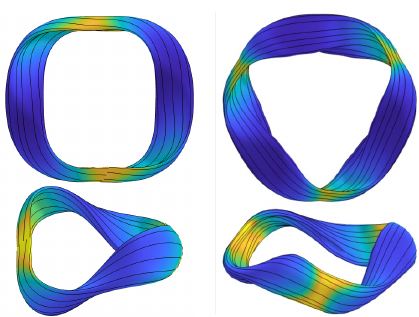}
    \caption{Two views of the plasma boundaries (top and bottom) of two compact $A=6.7$ SQuIDs with $n_\text{fp}=2$ (left) and $n_\text{fp}=3$ (right).}
    \label{fig:3Dplots2}
\end{figure}

Our findings suggest significant flexibility in the design of reactor-relevant QI stellarators, opening up the possibility to explore the vast parameter space of SQuIDs.  In \cref{fig:3Dplots2}, we present two additional SQuID configurations with aspect ratio $A=6.7$, one with three field periods and the other with two field periods.
These SQuIDs have excellent QI quality and negligible fast-ion losses at reactor scales, and their compactness is an attractive feature in an economical fusion reactor.
The $\iota$ profiles avoid low-order rational surfaces, but approach such a surface at the boundary, which should yield a chain of magnetic islands, compatible with an island divertor.

The SQuID with $n_\textrm{fp}=2$ was additionally optimized for compatibility with filamentary coils in addition to all other properties-of-interest, using the figure-of-merit described by \citet{kappel2023magnetic}.
We expect similar or, perhaps, lower ITG turbulent-driven heat losses than the $n_\textrm{fp}=4$ SQuID presented, according to values obtained for the target $f_{\nabla s}$, in combination with the smaller aspect ratios \cite{navarro2023assessing}.
These examples illustrate the reliability of our new target function, as various different optimization criteria can be ``baked in'' to SQuIDs with relative ease.

In this work, we focused on aspects of QI stellarator optimization related to plasma physics, but engineering issues also must be considered.
Further exploration of SQuID diversity, including coil optimization and system studies aimed at a stellarator reactor solution, is currently underway.


\begin{acknowledgments}
The authors thank J. Geiger, P. Costello, R. Mackenbach, and D. Dudt for their helpful comments.
We also thank D. Spong for providing the QPS equilibrium, and E. S{\'a}nchez, I. Calvo, and J. L. Velasco for the CIEMAT-QI equilibrium. 
Optimizations and simulations were performed at the Max Planck Computing and Data Facility (Germany) and the Cineca HPC (Italy). A. Goodman is supported by a grant from the Simons Foundation (560651). 
This work has been carried out within the framework of the EUROfusion Consortium, funded by the European Union via the Euratom Research and Training Programme (Grant Agreement No. 101052200 - EUROfusion). The views and opinions expressed are those of the authors only, and do not necessarily reflect those of the E.U. or the European Commission. 
\end{acknowledgments}


%

\end{document}